\documentclass{aa}

\usepackage{graphics}

\begin{document}

\title{Line formation regions of the UV spectrum of {\object CI Cygni}}

\author{ J. Miko{\l}ajewska \inst{1} \and M. Friedjung \inst{2} \and C. Quiroga\inst{3,4}} 

\offprints{J. Miko{\l}ajewska} 
\mail{mikolaj@camk.edu.pl}

\institute{
 Nicolaus Copernicus Astronomical Center,
 Bartycka 18, 00-716 Warsaw, Poland
\and
Institut d'Astrophysique de Paris - UMR 7095, CNRS/Universite Pierre et Marie
Curie,
98 bis Boulevard Arago, 75014 Paris, France
\and
Consejo Nacional de Investigaciones Cient\'{\i}ficas y T\'ecnicas de la
Rep\'ublica Argentina (CONICET)
\and
   Facultad de Ciencias Astron\'omicas y Geof\'{\i}sicas - Universidad Nacional de La Plata
- La Plata - Argentina}

\date{Received , Accepted...}

\titlerunning{UV spectroscopy of {\object CI Cyg}}
\authorrunning{Miko{\l}ajewska et.al.}

 \abstract
   {}
   {They are the interpretation of the emission line formation regions in
{\object CI Cygni}}
   {They involve the examination of radial velocities and fluxes of
ultraviolet emission lines at different 
epochs, deduced from archival IUE and GRHS/HST spectra.}
   {The line fluxes give electron 
densities and were in addition used to calculate emission measures, suggesting line formation in regions rather smaller than the binary separation.
Examination of the radial velocities led to us to find a systematic redshift of  the high ionization resonance lines with respect
to the intercombination, and \ion{He}{ii} lines.
Possible explanations of the redshift and the high resolution GHRS \ion{C}{iv} profile are discussed. We favour that involving resonance line absorption by a circum-binary region  most  probably in an asymmetric wind interaction shell or in a wind from the accretion disk. 
    }
   {}

\keywords{Stars: binaries: symbiotic -- Stars: fundamental parameters --
  Stars: mass loss -- Stars: individual: \mbox{\object CI Cyg}}

\maketitle

\section{Introduction}

{\object CI Cyg} is a symbiotic binary, containing a cool giant, of type M5.5 (M\"urset \&
Schmid \cite{ms99}), and a much more compact companion. The infrared spectrum shows
no sign of circumstellar dust. It is moreover particularly fruitful to study
{\object CI Cyg}, because it undergoes eclipses. A detailed investigation of this object
was performed by Kenyon et al. (\cite{ken91}). They concluded that the compact object
was a main sequence star of $0.5\, M_{\sun}$, surrounded by a large accretion
disk. Recent work shows however that the compact component of this binary may 
not be a main sequence star (e.g. Miko{\l}ajewska \cite{jm03}).

{\object CI Cyg} is one of the very few symbiotic binaries in which the cool giant fills
or nearly fills its Roche lobe (e.g. Miko{\l}ajewska \cite{mik01}, and references therein).
{\object CI Cyg} is also one of the very few symbiotic systems for which a
high quality radio spectrum, covering the range between 6 cm and 850 $\mu$m,  
has been obtained (Miko{\l}ajewska \& Ivison (\cite{mi01}).
These radio data enabled them to determine the free-free turnover 
frequency of the ionized component, and to critically test the known
models for radio emission from symbiotic stars.
Unfortunately, they ruled out the two most popular models: 
ionization of the red giant wind by Lyman continuum photons from
its hot companion, and emission resulting from 
the interaction of winds from the two binary components.

\begin{table}
\caption[]{Log of the UV observations of {\object CI Cyg}}
\label{obslog}
\begin{tabular}[bottom]{lcccc}
\hline
\noalign{\smallskip}
Image No. & Date & MJD & Phase$^1$ & T$_{\rm exp}$ [sec]\\
\noalign{\smallskip}
\hline
SWP\,06818 &  Dec 11, 1979 & 44157 & 0.718  & ~5400\\
SWP\,09256 & Jun 6, 1980 & 44401 & 0.003 & 10800 \\
SWP\,13795 & Apr 24, 1981 & 44719 & 0.375 & ~4800\\
SWP\,14754 &  Aug 14, 1981 & 44831 & 0.506 & ~5400\\
SWP\,18602 &  Nov 20, 1982 & 45294 & 0.046 & 40800 \\
SWP\,37047 &  Sep 17, 1989 & 47786 & 0.959 & ~6480 \\
SWP\,45102 &  Jul 9, 1992 & 48812 & 0.159 & ~5400 \\
SWP\,50760 &  May 14, 1994 & 49486 & 0.946 & ~6300\\
SWP\,51109 &  Jun 16, 1994 & 49520 & 0.986 & 23640 \\
SWP\,52838 &  Nov 20, 1994 & 49676 & 0.168 & 16200 \\
SWP\,54471 &  Apr 20, 1995 & 49827 & 0.345 & 21600 \\
SWP\,55894 &   Sep 10, 1995& 49970 & 0.513 & 18000 \\
\hline
Z1705107M &  Oct 11, 1993 & 49272 & 0.696 & 1795  \\
Z1705109T &  Oct 11, 1993 & 49272 & 0.696 & ~~707  \\
Z170510BT &  Oct 11, 1993 & 49272 & 0.696 & ~~816  \\
Z170510EM &  Oct 11, 1993 & 49272 & 0.696 & ~~816  \\
\noalign{\smallskip}
\hline
\end{tabular}

\noindent $^1$  Phase calculated from 
${\rm Min}(m_{\rm pg}/B)={\rm JD}\,2442687.1 + 855.6 \times E$
(Belczy{\'n}ski et al. \cite{bel00}). 
 \end{table}

A systematic shift in radial velocity between ultraviolet intercombination
lines and ultraviolet high ionization permitted resonance emission lines in
symbiotic binary spectra, was found by Friedjung et al. (\cite{fsv83}).
The latter are generally redshifted with respect to the former, and  
Miko{\l}ajewska \& Friedjung ({\cite{mf05}) have recently reported 
an orbital variation of this redshift in {\object CI Cyg}.
Unlike the intercombination
lines, the resonance lines are expected to be optically very thick. Their
redshift might be due to the absorption of both the continuum and part of the
emission line by P Cygni profile absorpion components, or only to a radiative
transfer effect produced by photon scattering in an expanding medium. 

In the present work, we have examined in detail the effect for the
symbiotic binary, {\object CI Cyg}, and we find some support for the P Cygni profile
interpretation. 
Our data base is presented in  Sect.2, analyzed in Sect.3, and the results discussed in Sect. 4.  We conclude with a brief summary of these results in
Sect. 5.

\begin{table}
\caption[]{Emission line fluxes and radial velocities from GHRS spectra
(MJD=49271/$\phi=0.696$)}
\label{hst}
\begin{tabular}[bottom]{lcc}
\hline
\noalign{\smallskip}
Line ID & Flux$^1$ & $v_{\rm rad}$ [km\,s$^{-1}$] \cr
\noalign{\smallskip}
\hline
\ion{Si}{iv}\, 1393.755 & ~28 &  34 \cr
\ion{Si}{iv}\,1402.769 & ~19 & 29 \cr
\ion{O}{iv}]\,1397.166 & ~~3 & 35 \cr
\ion{O}{iv}]\, 1399.731 & ~~7 & 25 \cr
 \ion{O}{iv}]\,1401.115 & ~36 & 26 \cr
\ion{O}{iv}]\,1404.740 & ~15 & 33\cr
\ion{O}{iv}]\,1407.333 & ~~9 & 26 \cr
\ion{S}{iv}]\,1406.004 & ~~5& 23 \cr
\ion{S}{iv}]\,1416.872 & ~~4 & 20 \cr
\ion{C}{iv}\,1548.202 & 350 & 32 \cr
\ion{C}{iv}\,1550.774 & 234 & 30 \cr
\ion{He}{ii}\,1640.470 & 205 & 12 \cr
\ion{O}{i}]\,1641.310 & ~~6 & 14 \cr
\ion{O}{iii}]\,1660.800 & ~25 & 20 \cr
\ion{O}{iii}]\,1666.150 & ~68 & 18 \cr
\ion{Mg}{v}\,2783.03 & 21 & 89 \cr
\ion{Mg}{ii}\,2796.352 &~30 & ~39 \cr
\ion{Mg}{ii}\,2803.531 & ~22 & ~35 \cr
\noalign{\smallskip}
\hline
\end{tabular}

\noindent $^1$  $10^{-14}\, \rm erg\,s^{-1}cm^{-2}$. 
\end{table}

\section{Observations}

The log of observations used is shown in Table~\ref{obslog}. They include high spectral
resolution observations made with the IUE satellite, archived in the INES data
base and higher resolution GHRS/HST observations made on 1993 October, 11 in the
regions of different lines. Though the latter were made before the COSTAR
correction for the spherical aberration of the mirror, the effect should not be
large, as the Small Science Aperture was used (0.25''). Profiles are shown in
Fig.~\ref{profiles}. Radial velocities and fluxes of different lines were obtained by 
a Gaussian fit of the line profile
(Tables~\ref{hst},  ~\ref{fluxes}, and ~\ref{rvel_tab}).
Radial velocity curves are shown in Fig.~\ref{rvel}.

\begin{figure}
 \resizebox{\hsize}{!}{\includegraphics{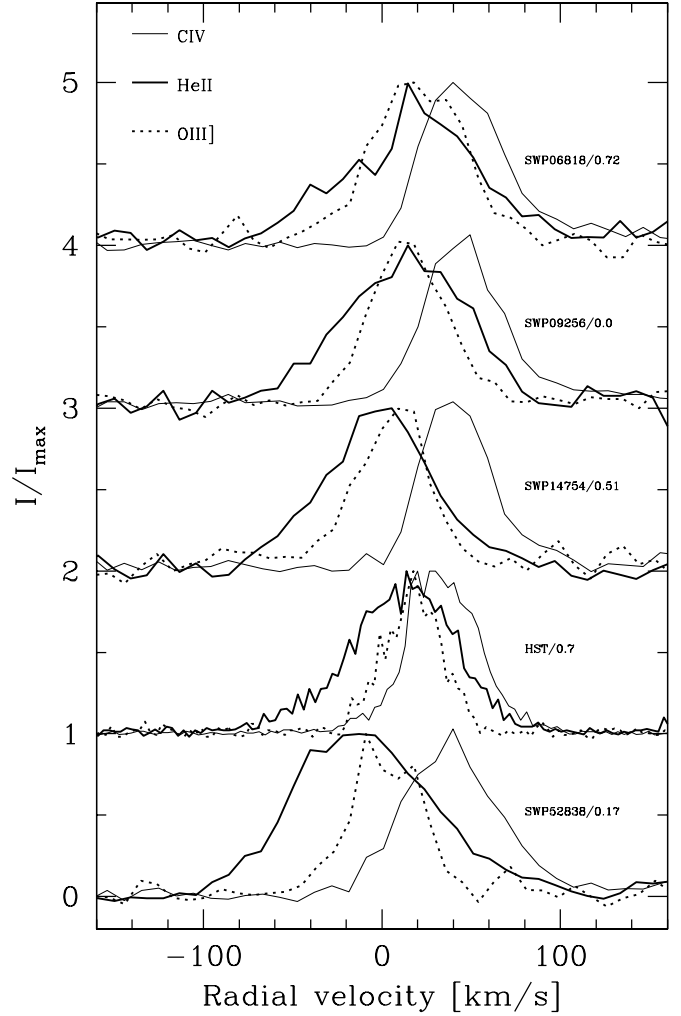}}
 \caption{Examples of emission line profiles of \ion{He}{ii}\,1640~\AA\, \ion{C}{iv}\,1548~\AA\
and \ion{O}{iii}]\,1666~\AA\ in {\object CI Cyg}. The numbers on the right side refer to the number of the spectrum and orbital phase, respectively.}
  \label{profiles}
\end{figure}

\begin{table*}[bht]
\caption[]{Emission line fluxes from IUE high resolution spectra 
(in units of $10^{-14}\, \rm erg\,s^{-1}cm^{-2}$)}
\label{fluxes}
\begin{tabular}{cccccccccccc}
\hline 
MJD & $\phi$ & \ion{N}{v} & \ion{Si}{iv} & \ion{O}{iv}] & \ion{N}{iv}] &
\ion{C}{iv}& \ion{He}{ii} & \ion{O}{iii}] & \ion{N}{iii}] & \ion{Si}{iii}] & \ion{C}{iii}]  \cr 
  &  & 1238/42 & 1394/1403 & 1400/01/05 & 1486 & 1548/50 & 1640 & 1660/66 &
1749/50/52/54 & 1892 & 1909 \cr
\hline
44157 & 0.718 & 100/37: & 63/52 & 40/.../21 & 180 & 800/588 & 336 & 35/123 & 
40/133/41/... & 248 & 577s \cr
44401 & 0.003 & 25/... & 48/44 & .../.../... & 82 & 489/345 & 137 & 23/156 & 
18/76/41/...  & 90 & S \cr
44719 & 0.375 & 68:/... & 37:/42: & .../.../... & 180 & 410/271 & 336 & 45/197 &
30/65/.../... & 99 & 331 \cr
44831 & 0.506 & 64/41 & 41/... & .../.../... & 166 & 350/249 & 360 & 54/185 & .../64/.../... & 127 & 443 \cr
45294 & 0.046 & 39/21 & 17/12 & .../.../... & 50 & S/S & S & 16/71s & .../32/.../... & S & S \cr
47786 & 0.959 & .../... & .../... & .../.../... & 39 & 236/212 & 289 & 6/50 & .../18/.../22 & 37 & 114 \cr
48812 & 0.159 & 50:/30: & .../... & .../.../... & 135 & 266/189 & 557 & 54/123 & .../65/.../... & 62 & 209 \cr
49486 & 0.946 & 36:/31: & 38:/21: & .../.../... & 65 & 401/350 & 419 & .../60 & .../38/.../... & 33 & 155 \cr
49520 & 0.986 & 59/... & 28/... & .../.../16 & 52 & S/268 & 300s & 5:/67 & .../16/.../... & 26 & 110s \cr
49676 & 0.168 & 100/64 & 26:/11: & 17/.../23 & 175 & 327/249 & 583 & 40/142 & 12/47/12/20 & 58 & 200s \cr
49827 & 0.345 & 332/178 & 99/61 & 11/95/35 & 221s & S/S & 480s & 82/220s & 28/67/13/12 & S & S \cr
49970 & 0.513 & 389/205 & 120/69 & 22/.../55 & 279 & S/S & 620s & 115/304 &
28/82/22/22 & 110s & S \cr
\hline \end{tabular}

\noindent s -- a few pixels saturated; S -- saturated; "..." -- too noisy; ":" -- uncertain.
\end{table*}

\begin{table*}[bht]
\caption[]{Radial velocities of the  IUE emission lines (in units of km\,s$^{-1}$).}
\label{rvel_tab}
\begin{tabular}{cccccccccccc}
\hline 
MJD & $\phi$ & \ion{N}{v} & \ion{Si}{iv} & \ion{O}{iv}] & \ion{N}{iv}] &
\ion{C}{iv}& \ion{He}{ii} & \ion{O}{iii}] & \ion{N}{iii}] & \ion{Si}{iii}] & \ion{C}{iii}]  \cr 
  &  & 1238/42 & 1394/1403 & 1400/01/05 & 1486 & 1548/50 & 1640 & 1660/66 &
1749/50/52/54 & 1892 & 1909 \cr
\hline
44157 & 0.718 & 37/35: & 36/42 & 39/.../44 & 26 & 43/42 & 19 & 20/20 & 19/14/8/... & 21 & {\it 15} \cr
44401 & 0.003 & 47/... & 38/41 & .../.../... & 19 & 44/40 & 14 & 26/14 & 
9/10/13/...  & 12 & {\it 10} \cr
44719 & 0.375 & 27:/... & 36:/32: & .../.../... & ~8 & 39/35 & -6 & 6/7 &
$-9:$/10/.../... & 11 & ~2 \cr
44831 & 0.506 & 19/22 & 26/... & .../.../... &  ~4 &  40/39  & -2 & 11/7  & .../ 4/.../... & 10 & 4 \cr
45294 & 0.046 & 27/20 & 23/27 & .../.../... & ~2 & {\it 38/29} & {\it 2} & -4/{\it 0} & .../-8/.../... & {\it -5} & {\it -1} \cr
47786 & 0.959 & .../... & .../... & .../.../... & ~1 & 27/25 & ~3 & 3/3 & .../3/.../5 & ~6 & ~4 \cr
48812 & 0.159 & 21:/21: & .../... & .../.../... & -3 &  33/ 34 & -17 & 3/3  & .../ 1/.../... & ~8 & ~0 \cr
49486 & 0.946 & 15:/24:  & ~9:/27: & .../.../... & ~7 & 32/29 & 10 & .../2 & .../4/.../... & ~1 & ~9 \cr
49520 & 0.986 & 21/... & 22/... & .../.../12 & -3 & {\it 29}/22 & {\it ~2} & .../-8 & .../-3/.../... & -4 & {\it -2} \cr
49676 & 0.168 & 33/31 & 42/35 & -1/.../4 & ~2 & 35/35 & -12 & 1/2 & 4/3/-1/7 & ~7&{\it ~3} \cr
49827 & 0.345 & 16/17 & 25/19 & 2/7/13 & ~2 & {\it 31/27} & {\it -23} & 0/6 & 6/4/0/1 & {\it  ~8} & {\it  ~3} \cr
49970 & 0.513 & 33/29 & 35/31 & 14/.../31 & 17 & {\it 39/34} & {\it -4} & 18/16 &
17/21/11/19 & {\it 25} & {\it 22}\cr
\hline \end{tabular}

\noindent "..." -- too noisy; ":" -- uncertain; for saturated lines the radial velocities (itallic) were obtained by fitting their wings.
\end{table*}

\section{Results and interpretation}

\subsection{Radial velocity curves and emission line flux changes}

\ion{He}{ii} traces the motion of the hot component but the systemic
velocity, $\gamma = 0.6 \pm 1.7$ km\,s$^{-1}$,
seems to be blueshifted by about 18 km\,s$^{-1}$ with respect
to the M giant orbital solution of Kenyon et al. (\cite{ken91}). A similar
blueshift in the 
systemic velocity was found for the radial velocity curve for the 
optical \ion{He}{ii}\,4686\,\AA\ line (Kenyon et al. \cite{ken91}).
The amplitude of the circular orbit, $K_{\rm h} = 17.3 \pm 2.8$ km\,s$^{-1}$,  combined
with that of the M giant, $K_{\rm g} = 6.7 \pm 0.8$ km\,s$^{-1}$
implies  that $q = M_{\rm g}/M_{\rm h} = 2.58^{0.82}_{0.65}$,
and is consistent with other estimates (see Kenyon et al. \cite{ken91}
for details). We should note in addition that \ion{He}{ii} shows no
systematic shift in radial velocity between decline and quiescence.

\begin{figure}
 \resizebox{\hsize}{!}{\includegraphics{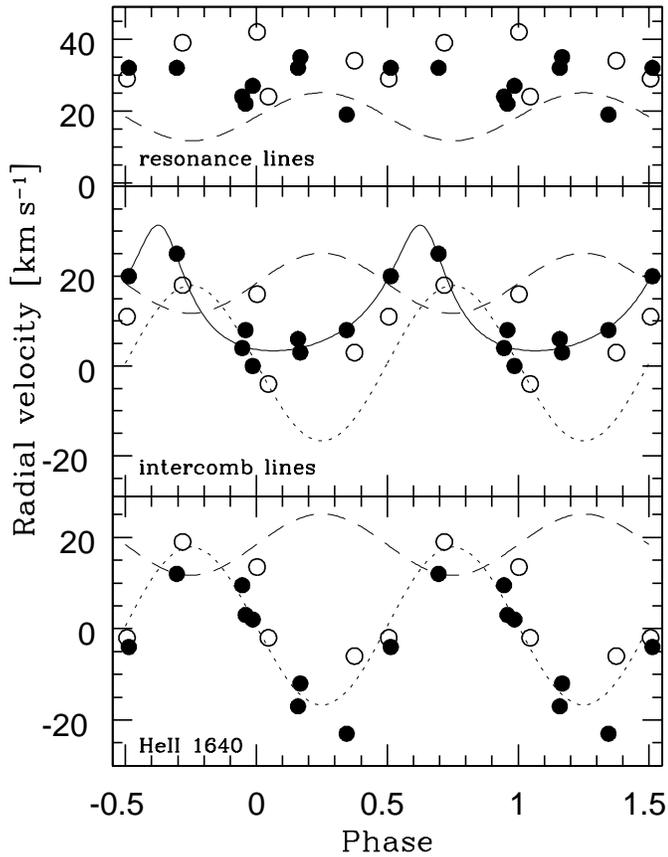}}
 \caption{Radial velocity curves for {\object CI Cyg}. The dashed line repeats the 
orbit of the M giant and the dotted line -- the circular orbit solution for 
the \ion{He}{ii}\,1640\,\AA\ emission line, respectively. The solid line corresponds to the elliptical orbit solution for the intercombination line (see text for details).
Open and closed symbols correspond to the late decline (1979-82) and 
quiescent data (after 1983), respectively.}
  \label{rvel}
\end{figure}

The intercombination lines are roughly in phase with the hot component.
However the amplitude, $K \sim 10$ km\,s$^{-1}$ is much lower,
the systemic velocity, $\sim 10$ km\,s$^{-1}$, is blueshifted with respect
to the systemic velocity of the red giant, $\gamma_\mathrm{g} = 18.4 \pm 0.4$
km\,s$^{-1}$ (Kenyon et al. \cite{ken91}), and 
the orbit seems to be eccentric with
$e \sim 0.4$, the longitude of periastron, $\omega \sim 0^{\degr}$, and the
spectroscopic conjunction occurs at $\phi \sim 0.8$.
In fact, they follow with very good precision the orbit found
for the optical \ion{Fe}{ii} emission lines by  Kenyon et al. (\cite{ken91}).
The solid line in the middle panel of Fig.~\ref{rvel} corresponds to an
orbital solution with the same parameters as those for \ion{Fe}{ii} in Table 8
of Kenyon et al. (\cite{ken91}) except for the amplitude and the systemic
velocity which
were fitted with our quiescent data. There is also an 
interesting systematic difference between the radial 
velocity curve formed by the quiescent (after 1983), and late decline
(1979--82) data, respectively (Fig.~\ref{rvel}): although both data sets seem
to follow the same orbit  the data from the decline 
are slightly blueshifted (by $\sim 7 \pm 2$ km\,s$^{-1}$) with respect to the quiescent data. The only point above the solid curve in Fig.~\ref{rvel} is
the measurement from the spectrum obtained in the midle of the 1980
eclipse. We note that the radial velocities for the other emission lines
derived from this spectrum are redshifted by $\sim 20$ km\,s$^{-1}$ with respect to
the data obtained near the other eclipses, and it may be due to uncertainty in
the wavelength calibration.

The blueshift of the systemic velocity of both the optically thin
intercombination lines
and the usually not optically very thick \ion{He}{ii}\,1640\,\AA\  line
suggests  line formation  regions approaching the observer whereas 
the systemic velocity difference between the intercombination lines and
\ion{He}{ii}\,1640\,\AA, appears to
indicate a velocity stratification and/or acceleration  effects in the ionized
nebula, with
electron temperature variations perhaps affecting collisionally excited lines. Probably for the same reason,  the \ion{He}{ii} line shows  a systematically  broader profile than the intercombination \ion{O}{iii} line  (Fig.~\ref{profiles}).

The radial velocities of the optically thick resonance lines are almost stationary, and redshifted by $\sim$ 10$-$20 km\,s$^{-1}$ with respect to the systemic velocity of the red giant. 
Miko{\l}ajewska \& Friedjung (\cite{mf05}) showed that 
the velocity difference between the resonance and intercombination
emission lines seems to follow the radial velocity of the cool
giant, with  a comparable amplitude, $\sim 6$ km\,s$^{-1}$, and average
velocity,
$\sim 21$ km\,s$^{-1}$, almost the same as the systemic velocity. 
We note however that this effect may not be physical because of
the apparent constancy of the resonance line radial velocity; which means
that the radial velocity difference curve is a mirror image of the radial
velocity curve for the intercombination lines.

A comparison of the \ion{C}{iv} emission line profiles with those of \ion{He}{ii} ( Fig.~\ref{profiles}), and particularly the best resolution profiles taken with the HST (Fig.~\ref{hst_profiles}) shows that the red wings of both lines overlap whereas the blue wing of the \ion{C}{iv} line seems to be missing, possibly due to the presence of a blue shifted absorption
component in a P Cygni profile. The corrected emission component might in
that case trace the  motion of the hot component. A P Cygni profile of
this sort seems to be mostly related to the
hot  component's activity: it was certainly more pronounced 
in the early 1980's (decline from the 1975 outburst) than in the
1990's when the system had reached the quiescent state. 
A weak absorption component seems to be also present in the blue wing of
the \ion{He}{ii} profile taken in 1979 ($\phi \sim 0.7$) whereas it may be 
marginally visible in the HST/GHRS profile taken in 1993.

It seems that some radial velocities are related to  the level of
ionization:
the HST data give radial velocities of 29 km\,s$^{-1}$ for the
intercombination \ion{O}{iv}], 13 km\,s$^{-1}$ for the intercombination
\ion{O}{iii}] and 11 km\,s$^{-1}$ for the \ion{O}{i}] lines. However, there
is no systematic difference  in the velocities of \ion{C}{iv} and \ion{Si}{iv}.

\begin{figure}
 \resizebox{\hsize}{!}{\includegraphics{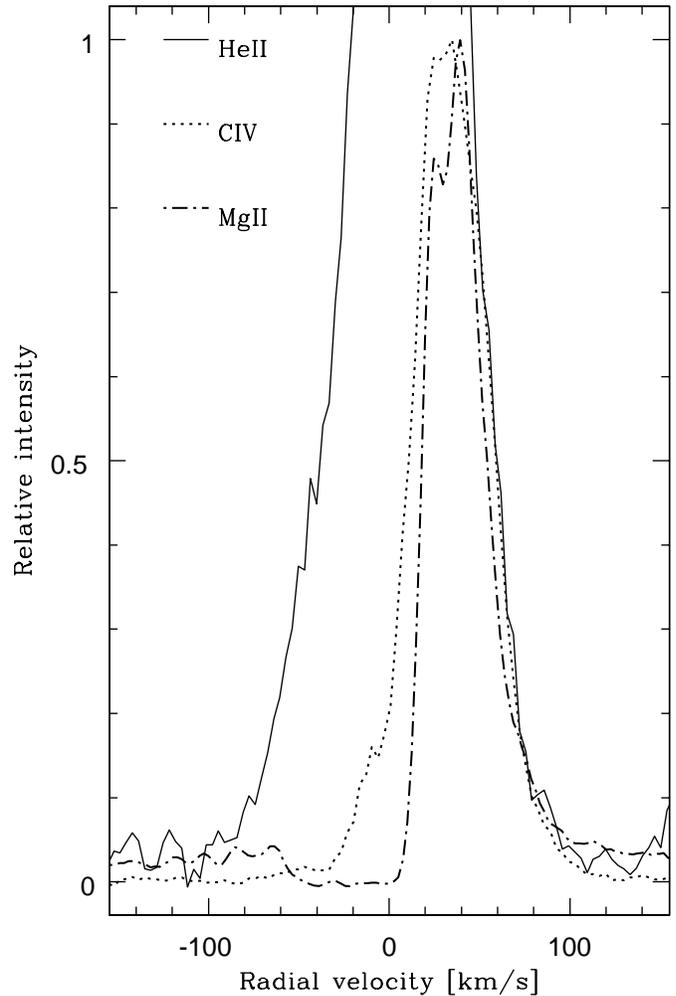}}
 \caption{The smoothened and normalized \ion{He}{ii}\,1640 (solid line), 
average \ion{C}{iv} profiles (dotted line) and \ion{Mg}{ii} (dashed line) profiles, observed with the HST/GHRS. The \ion{He}{ii} profile was increased by a factor of about 2, 
in order to fit the red wing to the \ion{C}{iv} and \ion{Mg}{ii} profiles.
}
  \label{hst_profiles}
\end{figure}

The resonance line ratios indicate moderate optical depth effects.
In particular, the average ratios are: 
$F(\ion{N}{v}\,1238/1240)=1.9 \pm 0.3$, 
$F(\ion{Si}{iv}\,1394/1403)=1.7 \pm 0.2$ (1.5 for HST spectra), and
$F(\ion{C}{iv}\,1548/1550)=1.33 \pm 0.05$ (1.5 for HST spectra),
respectively. These ratios do not seem to vary with the orbital 
phase nor the activity, although the line fluxes show both orbitally 
related and secular changes (Fig.~\ref{lcurves}).
In particular, the permitted resonance and \ion{He}{ii} 1640 line fluxes were increasing as the system declined from
the last outburst, and then stabilized at some high level whereas the
intercombination line fluxes tended to decrease.
Both the permitted and intercombination line fluxes change with the orbital phase, 
and the minima in all lines but \ion{He}{ii} are much deeper and broader in quiescence than
during the decline in activity. 
The same effect is present for the optical Balmer
lines.

\begin{figure}
 \resizebox{\hsize}{!}{\includegraphics{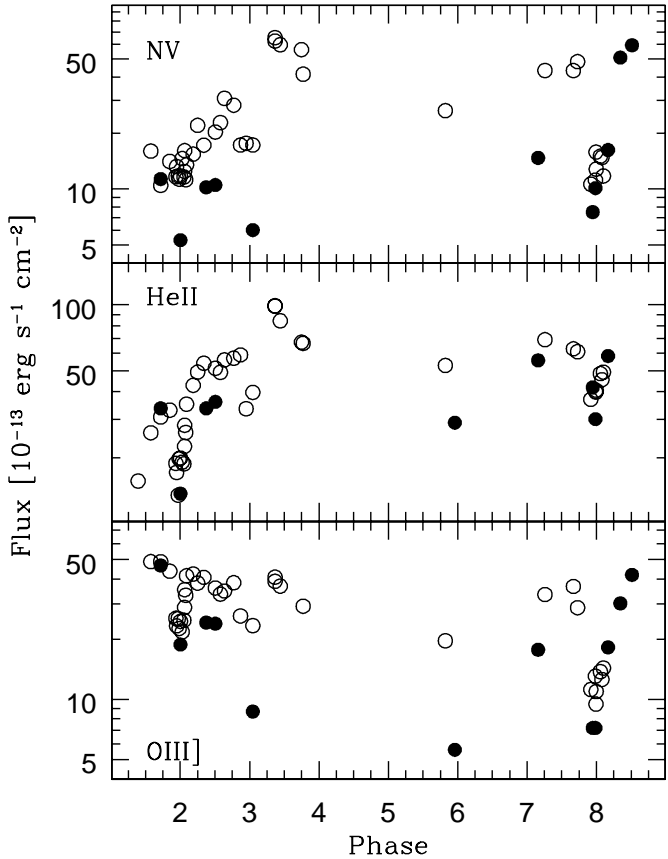}}
 \caption{Variation of line fluxes for \ion{N}{v}\,1240,  \ion{He}{ii}\,1640
and \ion{O}{iii}]\,1664 emission lines. The open and filled circles correspond to the data from the high and low resolution IUE spectra, respectively.}
  \label{lcurves}
\end{figure}

The observed orbitally related changes of the emission line fluxes require the main line formation regions to be located inside the binary system, with possible stratification.  
The fact that the resonance lines show practically the same secular trend and eclipse behaviour as the \ion{He}{ii} line suggests that they are formed in the same region, and that their profiles should be intrinsically the same. Thus the lack of orbitally related radial velocity changes together with their redshift with respect to the \ion{He}{ii} and intercombination lines indicates that the resonance line profile must be seriously affected by an absorption component.  Moreover, the emission must be absorbed outside the binary system in very extended region. 
The \ion{He}{ii} 1640\,\AA\ line would be not affected by this circumstellar absorption because it is the recombination line from a high excitation level (41 eV).
   
The \ion{O}{iii}] line ratio, the average 
$F(1666/1660)=3.85\pm0.45$,
is always bigger than the optically
thin value, and the effect is much stronger during the eclipse
of the hot component. The effect was also stronger before 1983 
when {\object CI Cyg} was still declining from the large 1975 outburst.
If this anomalous line ratio is due to
Bowen pumping of optically thick lines 
as proposed by Kastner et al. (\cite{kastner89}), the pumping must
be more effective during the decline as well as in the region(s) 
visible during the eclipse.
The fact that the out-of-the-eclipse intensity 
of the Bowen lines 3133\,\AA\ and 2837\,\AA\ decreased after 
1983 whereas the intercombination doublet itself remained practically 
constant, suggests that the Bowen excitation rate was indeed higher before
1983 (Kenyon et al. \cite{ken91}) than in the 1990's. 
On the other hand, both the Bowen lines and the intercombination doublet 
show an eclipse effect, although the effect is less pronounced for 
the Bowen lines.

\subsection{Emission measures}

The density sensitive intercombination line ratios measured on IUE spectra
indicate
$n_\mathrm{e} \sim 4 \times 10^9$ and $\sim 3\times 10^9\, \mathrm{cm}^{-3}$ 
for \ion{N}{iii}] and \ion{O}{iv}] (1404.8:1401.2), 
respectively (Nussbaumer \& Storey \cite{ns79}; Nussbaumer \& Storey \cite{ns82}). 
We note, however, that the \ion{O}{iv}]\,1404.8 line 
is blended with \ion{S}{iv}] line, which contributes  $\sim 10\,\%$
of the total line flux.
The \ion{O}{iv}] line ratios measured on the HST spectrum indicate
$n_\mathrm{e} \sim 2 \times 10^9$ (1399.8:1401.2:1404.8) 
and $\la 10^{10}\, \mathrm{cm}^{-3}$ (1401.2:1407.4), respectively
(Harper et al. \cite{harper99}; Keenan et al. \cite{keenan02}).
Similarly, the \ion{S}{iv}]\,1416.9/1406.0 ratio indicates 
$n_\mathrm{e} \la \mathrm{a\,few} \times 10^9\, \mathrm{cm}^{-3}$
(Harper et al. \cite{harper99}; Keenan et al. \cite{keenan02}).
These electron densities are comparable to those derived for the \ion{He}{i}, 
Balmer \ion{H}{i}, and \ion{O}{iii}] emission region(s) by Kenyon et al. (\cite{ken91}). 
They may, however, not be representative of the resonance
line emission region.

Let us examine the sizes of the regions of emission of the ultraviolet high
ionization permitted zero volt lines of \ion{C}{iv}, \ion{N}{v} and \ion{Si}{iv}. These lines are
expected to be excited by electron collisions.
Assuming that the line emission is effectively optically thin, 
we can express the line luminosity as (e.g. Brown \& Jordan \cite{bj81}, Miko{\l}ajewska et al. \cite{jm88}):
\begin{equation}
4 \pi d^2 F_{\lambda, \mathrm{obs}} = 8.6 \times 10^{-6}  n_\mathrm{e}{\Omega_\mathrm{mn} \over g_\mathrm{m}} {N_\mathrm{E} \over N_\mathrm{H}}
{N_\mathrm{H} \over n_\mathrm{e}} 
\int\limits_\mathrm{V} {g(T) n_\mathrm{e}^2} dV
\label{emflux}\end{equation}
with 
\[
g(T)=T_\mathrm{e}^{-1/2} {N_\mathrm{ion} \over N_\mathrm{E}} \exp{-W_\mathrm{mn} \over k T_\mathrm{e}}.
\]
Here $d$ is the distance, $F_{\lambda, \mathrm{obs}}$ -- the reddening-corrected observed line flux,
$T_\mathrm{e}$ -- the electron temperature,
$N_\mathrm{ion}$,  $N_\mathrm{E}$, $N_\mathrm{H}$ and $n_\mathrm{e}$
are the number densities of the ion in question, of an element $E$, of hydrogen, and electrons, respectively,
$\Omega_\mathrm{mn}$ is the collision strength of the transition from level $m$ to
level $n$, $g_\mathrm{m}$ is the statistical weight of the lower level, 
and $W_\mathrm{mn}$ --  the excitation energy.

\begin{table}
\caption[]{Emission measures, $EM$, in units of $10^{57}\, \mathrm{cm}^{-3}$.}
\label{em}
\begin{tabular}[bottom]{cccccc}
\hline
\noalign{\smallskip}
MJD & Phase & \ion{N}{v} & \ion{Si}{iv} & \ion{C}{iv} & Note \cr
\noalign{\smallskip}
\hline
44831 & 0.506 & 140 & 31 & 22 & (1) \cr
          &          & 2.8 & 2.6 & 1.8 & (2) \cr
49271 & 0.696 &      & 24 & 21 & (1) \cr
          &          &      & 2.0 & 1.7 & (2) \cr
\noalign{\smallskip}
\hline
\end{tabular}

\noindent (1)  $T_\mathrm{e}=15\,000\, \mathrm{K}$; (2)  $g(T)=0.7 g_\mathrm{max}(T)$.
\end{table}

The function $g(T)$ peaks strongly at a certain temperature, and it is often assumed 
$g(T)=0.7 g_\mathrm{max}(T)$ (Pottasch \cite{pot64}). 
Such an assumption is valid for the line formation in a hot region, like the solar transition region, e.g. an accretion disk corona. 
Table~\ref{em} gives emission measures, 
$EM=\int\limits_\mathrm{V} n_\mathrm{e}^2dV$
calculated from the observed emission line fluxes, corrected for $E_\mathrm{B-V}=0.4$ with the reddening curve of Seaton (\cite{se79}), and 
adopting $d=2$ kpc (Miko{\l}ajewska \& Ivision \cite{mi01}; Kenyon et al. \cite{ken91}), for two values of $g(T)$: $g(T)=0.7 g_\mathrm{max}(T)$ and $T_\mathrm{e}=15\,000$ K,  representative  of line formation in a hot shocked region and a photoionized region, respectively. We also 
assumed solar abundances (cf. Proga et al. \cite{proga96}), and supposed that 
most of the atoms of the
element in each line emission region were in the same state of ionization.

The present estimates take account of the fairly small contribution of  the
electrons due to ionization of helium, supposing it singly ionized in the
Si$^{+3}$ region and doubly ionized in the regions of C$^{+3}$ and of
N$^{+4}$. The collision strengths used are those tabulated by Brown and
Jordan (\cite{bj81}), and Miko{\l}ajewska et al. (\cite{jm88}). 
In view of the presence of optical depth effects, we calculated
the total rate of collisional excitation for each multiplet considered,
assuming that photons were not lost by other sources of absorption.

It is noteworthy that if $T_\mathrm{e} \sim 15\,000\, \mathrm{K}$ the emission 
measure is $EM \sim \mathrm{a\,\,few} \times 10^{58}\, \mathrm{cm}^{-3}$ 
for both the resonance lines and \ion{He}{ii}\,1640\, \AA. 
For a density $n_\mathrm{e} \sim 10^{10}\, \mathrm{cm}^{-3}$, typical for
the central, eclipsed nebular region(s) in {\object CI Cyg} (e.g. Kenyon et al. \cite{ken91}), 
 emission measures of this order imply spherical radii lower than 
the binary separation, in agreement with some obscuration effects 
clearly visible in \ion{N}{v} lines.

We may note that such values of the emission measure are similar to the
radio volume 
emission measure estimate of Miko{\l}ajewska and Ivison (\cite{mi01}), but in
view of the large length scale of the radio emission, of the order of $10^{15}$ cm (larger than
the binary separation), this is presumably a
coincidence.
The radio emission region, however, can be the same as  
the resonance line absorbing region, if the optical thickness of the \ion{C}{iv} lines is large.

\section{Discussion}
The fact the giant component of {\object CI Cyg} fills or nearly fills its tidal lobe implies that the usually adopted models for the symbiotic nebulae (with the main emission region in the spherically symmetric red giant wind ionised by the hot companion) cannot be a good match to the true conditions in the emission line region(s). The observed orbitally related changes of radial velocities and fluxes of most UV emission lines require the main formation region to be located  inside the binary system. It is, however, in the partially ionised mass-loss stream and the environment of the accretion disk and the hot component rather than in a hemispherical shell just outside the red giant wind photosphere. 

At the present stage, we cannot rule out completely radiative transfer as being responsible for the redshift of resonance lines with respect to the intercombination and \ion{He}{ii} lines. 
However a P Cygni profile explanation appears more likely, because of the lack of radial velocity changes of the resonance lines, suggesting that the lines are seriously affected by circumstellar absorption. \ion{He}{ii} 1640\,\AA\  line is optically much thinner, as its lower level is at 41 eV and is not affected.
As far as the resonance line emission is concerned, our emission measures suggest line formation in regions smaller than the binary separation. 
The origin of the absorption component is however not obvious.

It cannot, in general,  be
due to interstellar line absorption (see discussion
in Friedjung et al. \cite{fsv83}).
However in the case of {\object CI Cyg}, two other objects in the same direction of
the sky, show similar \ion{C}{iv} (as well as \ion{Mg}{ii}) absorption to that
suggested by the spectrum of {\object CI Cyg}. We see the absorption in high
resolution IUE spectra of the X-ray binary {\object Cygnus X-1} and the Wolf-Rayet star
{\object HD~190918}, which are within 4$^o$ from {\object CI Cyg}. This means that we cannot be
certain to what extent the suggested {\object CI Cyg} absorption is really circumstellar.
The absorption of {\object HD~190918}, spread out between very small negative radial
velocities and about $-$80 km\,s$^{-1}$, has apparently no connection with the
much more blueshifted absorption components, due to the wind of that star.
However the spectra of the two other objects do not have \ion{N}{v}
absorption. In fact the {\object CI Cyg} \ion{N}{v} shift is at most only  slightly less
than the mean for the 6 high ionization lines (2$-$7 km\,s$^{-1}$ on spectra where
all 6 lines could be measured). This means that the redshift cannot be due to
interstellar line absorption in the blue wings of the \ion{C}{iv} doublet.

The redshift  cannot  either be accounted by absorption by
an iron curtain because the effect is present in all lines,
including \ion{N}{v} which should not be affected, whereas it is missing
in the \ion{He}{ii}\,1640\,\AA\ line which should be affected by the Fe$^+$-curtain
(e.g. Shore and Aufdenberg \cite{sa93}).

Another possible effect is absorption of resonance line emission from
excited \ion{Fe}{ii} levels, which results in pumping of highly excited levels
of that ion and strong emission produced by cascades from these pumped levels.
The emission of the \ion{C}{iv}\ 1550\,\AA\ ~and \ion{N}{v}\,1242\,\AA\ ~lines is
absorbed; if this absorption is optically thick, the profiles and relative
fluxes of the lines in the same multiplet will be affected (Eriksson et al
\cite{e03a}). However, according to Eriksson et al (\cite{e03b}), no such
pumping occurs for {\object CI Cyg}. The deviation of the  \ion{C}{iv} multiplet flux ratio
from the optically thin value rather suggests self absorption. In any case the normalized profiles the \ion{C}{iv}~ resonance
multiplet lines are virtually identical, except for small noise.

The apparent absorption component should not occur in the same region as the emission. The absorbing region would need to be larger with a lower source function and would be presumably circumbinary.
Such a region would need to be mainly in expansion, but some of the apparent
absorption is to the red of the cool giant systemic velocity, as given by Kenyon et al (1991). This means that parts
of the region (inner parts) would need to be contracting towards the central
binary. This might be just the gas pressure expansion velocity of a dense
medium into a much lower density medium.
Such a region might be due to the interaction between winds from the two binary components.
Although Miko{\l}ajewska \& Ivison (\cite{mi01}) ruled out the interacting wind model (assuming spherically symmetric winds from both components)  of Kenny (1995; see also Kenny \& Taylor \cite{kt05})  as a possible mechanism for the extended radio emission from {\object CI Cyg}, we note that non spherically symmetric interacting winds are still possible. In particular, the mass loss from the giant is strongly concentrated in a stream which should result in the formation of a disk of material orbiting the companion. So, any wind from the companion should be bipolar rather than spherically symmetric. Moreover the disk itself can be also a source of a wind. The geometry of the mass flow(s) in {\object CI Cyg} thus differs significantly from the simple geometries assumed for the existing models.
The resonance line absorption can be then located in the same region as the radio emission, for example, in a swept-up shell where a low-velocity material lost from the giant is overtaken by a higher velocity wind from the hot companion and/or accretion disk.
 
Let us note that short wavelength absorption is more clearly visible
in ultraviolet spectra of other symbiotic binaries during outburst. A
1984 outburst spectrun of {\object Z And} shows two blueshifted components, one with
a velocity of around 75$-$100  km\,s$^{-1}$ and the other with an edge velocity of
230 km\,s$^{-1}$ in \ion{C}{iv}, while the lower velocity component is at least
visible in \ion{N}{v}. In particular, blueshifted absorption at 120 km
s$^{-1}$ in \ion{N}{v} and \ion{C}{iv} was reported by Fernandez-Castro et al
(\cite{fe95}) for {\object Z And} at maximum. In addition blue shifted absorption has
also been seen for \ion{N}{v} when {\object AG Dra was in outburst, the velocity being
larger when the star was brighter (see Viotti et al \cite{vi84}).

The suggested ultraviolet P Cygni absorption is however not blueshifted 
enough to be produced by the wind of a compact object 
(lower main sequence star or more compact). 
Another possibility is that it is produced by the wind of an
accretion disk around the compact component, which is slowly accelerated. 
Such a wind could have a much higher velocity than the observed radial
velocity, because the disk would have a large inclination to the line of sight
in this eclipsing system. Note that the blue shifted absorption components  are
probably not produced by a jet perpendicular to the orbital plane; 
any jet perpendicular to the orbital plane would be almost
perpendicular to the line of sight and not absorbing.

The \ion{He}{ii}  line blueshift is explainable if the emission is formed in an
expanding medium above
the central regions of an optically thick accretion disk
The similarity of the intercombination line orbit to that of the optical
\ion{Fe}{ii} orbit might be understandable if the \ion{Fe}{ii} lines were
optically thin and formed near the intercombination line region. Kenyon et al
(\cite{ken91}) explained both the apparent eccentricity of the orbit and the
phase displacement of the spectroscopic conjunction by the formation of
\ion{Fe}{ii} near the inner Lagrangian point in the gas stream and/or bright
spot where the matter stream impacts the outer accretion disk. The small
difference in radial velocity amplitudes and phase shifts between the
intercombination and \ion{Fe}{ii} lines can in principle be explained by some
stratification effects in the stream. The change between decline and
quiescence might be related to a change in the properties of the disk.

\section{Conclusions}

Archival ultraviolet spectra taken by IUE and at higher resolution at one
epoch by the GHRS/HST are studied by us. We examined radial velocities and line
fluxes as well as higher resolution HST line profiles, taken at one epoch.
The line fluxes were used to determine electron densities and emission
measures; line formation in regions rather smaller than the binary separation
being indicated.

The relative systemic velocity  shifts of the \ion{He}{ii} and
intercombination lines may be explicable by expansion, possibly in a medium above an optically thick accretion disk, with stratification. In this framework changes in disk properties might be responsible for changes in the intercombination line radial velocity between decline and quiescence.

A systematic redshift 
between the optically thick resonance lines on the
one hand and the optically thin intercombination lines and the usually not
optically very thick \ion{He}{ii}\,1640\,\AA\  line on the other hand, like that of 
Friedjung et al. (\cite{fsv83}), was confirmed by us. Two possible explanations still exist. We favour that involving  a non-classical P
Cygni profile due to large circum-binary absorbing region, which is mainly
expanding, but of which parts (perhaps inner parts) are contracting towards the
binary. This region is most  probably an asymmetric wind interaction shell or a wind from the accretion disk. Such a region could also produce the observed radio emission.
The other explanation of the redshift, as due to radiative transfer effects, cannot yet however be completely eliminated. 
More theoretical work on the production of such a line shift is still required.

\begin{acknowledgements}
This research has been partly supported by KBN grants 5P03D\,019\,20,
and 1P03D\,017\,27, and by the European Associated Laboratory "Astrophysics Poland-France".
It also made use of the NASA Astrophysics Data System
and SIMBAD database.
\end{acknowledgements}

\end{document}